\documentclass[fleqn,usenatbib]{mnras}

\usepackage{newtxtext,newtxmath}

\usepackage[T1]{fontenc}
\usepackage{ae,aecompl}

\usepackage{graphicx}	
\usepackage{amsmath}	
\usepackage{amssymb}	

\title[Microlensing and Photon Bunching]{Microlensing and Photon Bunching: \\ The impact of decoherence}

\author[Lewis \& Tuthill]{
Geraint F. Lewis\thanks{E-mail: geraint.lewis@sydney.edu.au} and
Peter Tuthill
\\
Sydney Institute for Astronomy, School of Physics, 
A28, The University of Sydney, NSW 2006, Australia
}

\date{Accepted XXX. Received YYY; in original form ZZZ}

\pubyear{2019}

\begin{document}
\label{firstpage}
\pagerange{\pageref{firstpage}--\pageref{lastpage}}
\maketitle

\begin{abstract}
Gravitational microlensing within the Galaxy offers the prospect of probing the details of distant stellar sources, as well as revealing the distribution of compact (and potentially non-luminous) masses along the line-of-sight. Recently, it has been suggested that additional constraints on the lensing properties can be determined through the measurement of the time delay between images through the correlation of the bunching of photon arrival times; an application of the Hanbury-Brown Twiss effect. In this paper, we revisit this analysis, examining the impact of decoherence of the radiation from a spatially extended source along the multiple paths to an observer. The result is that, for physically reasonable situations, such decoherence completely erases any correlation that could otherwise be used to measure the gravitational lensing time delay. Indeed, the divergent light paths traverse extremely long effective baselines at the lens plane, corresponding to extremes of angular resolving power well beyond those attainable with any terrestrial technologies; the drawback being that few conceivable celestial objects would be sufficiently compact with high enough surface brightness to yield usable signals.
\end{abstract}




\section{Introduction}
\label{sec:introduction}
Gravitational microlensing has proven to be a powerful probe of the properties of the Milky Way. Originally envisaged as an effective tool to search for compact dark matter in the Galactic Halo~\citep{1986ApJ...304....1P}, the deflection of light by stellar masses has become a workhorse in imaging distant stellar surfaces~\citep[c.f.][]{2002MNRAS.335..539H}, and revealing the presence of exoplanets through perturbations to the lensing magnification~\citep[c.f.][]{2012ARA&A..50..411G}.

During a microlensing event, the resultant sub-images spawned by the lens occupy a realm of separation on the scale of milliarcseconds, and so have traditionally been regarded as unresolvable with observations focused on changes in integrated light (magnification).
With the recent advent of long baseline optical interfereometers~\citep{2001A&A...375..701D}, these scales have now become observable and just beginning to be within sensitivity limits for the world's most capable instruments. Indeed the potential for progress employing this approach was recently demonstrated by  \citet{2019ApJ...871...70D} who achieved single-epoch observation of a microlensing event using the Gravity beam combiner at the VLTI interferometer, proceeding to show that useful constraints on the physical parameters of the event could be obtained from the data.

Entirely new strategies to obtain such constraints on the gravitational lensing configuration would be available if the time delay between the separate microlensing sub-images~\citep[c.f.][]{1964MNRAS.128..307R} could be measured.
While the light paths traversed by the separate components are widely separated in space, it turns out that the calculated time delays traversing the lensing system are actually quite small -- of order $\Delta t \sim 10^{-6}s$ -- a consequence of the geometrical and relativistic terms in the computation acting to almost cancel.
Although measuring such a small delay might appear difficult in practice, recently, \citet{2019MNRAS.486.5400S} proposed to use intrinsic correlations between photon arrival times as an embedded clock signal by which the two different light-paths might be synchronised and the delay between them obtained.
This is a proposed application of the Hanbury-Brown Twiss (HBT) effect, with \citet{2019MNRAS.486.5400S} suggesting that such photon bunching will be measurable for sources out to $10$kpc with 30m-class telescopes.
We note, in passing, that the essentials of this idea have resurfaced more than once; for example around a decade ago with the work of \citet{2008MNRAS.389..364B} which we briefly touch upon later in this manuscript.

In this paper, we revisit the analysis of photon bunching during gravitational microlensing events as framed by \citet{2019MNRAS.486.5400S}.
The structure of this paper is as follows: In Section~\ref{sec:theory}, we present the background, briefly outlining microlensing and photon bunching in Sections~\ref{sec:microlensing} and \ref{sec:photonbunch} respectively, including the argument given in~\citet{2019MNRAS.486.5400S}. In Section~\ref{sec:decoherence} we explore the impact of decoherence along the gravitationally lensed photon paths, demonstrating how this wipes out the prospect of detecting the expected time delay between lensed images. We present our conclusions in Section~\ref{sec:conclusions}.

\section{Background}
\label{sec:theory}

\subsection{Microlensing}
\label{sec:microlensing}
We will consider gravitational lensing of a point-like source by a point mass; for a more comprehensive review of gravitational lensing, the reader can consult \citet{1992grle.book.....S}. Following the notation of \citet{2019MNRAS.486.5400S}, a lens with a mass of $M$, with a source at a 
angular
position, $\beta$, results in a corresponding image at 
angular
position, $\theta$, given by the lensing equation
\begin{equation}
    \beta = \left( x - \frac{1}{x} \right) \theta_E
    \label{eqn:lensing}
\end{equation}
where $\theta_E$ is is the angular Einstein radius and is given by
\begin{equation}
    \theta_E^2 \equiv \frac{4 G M}{c^2} \frac{D_{LS}}{D_L D_S}
    \label{eqn:einstein}
\end{equation}
The point mass lens produces two images which are observed at 
angular
positions
\begin{equation}
    \theta_1 = x \theta_E\ \ \ {\rm and}\ \ \ \theta_2 = -\theta_E / x
    \label{eqn:images}
\end{equation}
where the images appear along the line on the sky connecting the lensing mass and the source.
Each image is magnified by
\begin{equation}
    A_1 = \frac{x^4}{x^4 - 1}\ \ \ {\rm and}\ \ \ A_2 = \frac{1}{x^4 - 1}
    \label{eqn:magnification}
\end{equation}
with the 
sum of 
absolute of
these being the total magnification of the source.

Photons travelling from the source to the observer in a gravitational lens travel a larger geometric distance than the direct path between the two, resulting in an longer travel time. However, there is an additional time delay effect due to the relativistic influence of the photon travelling through the gravitational potential of the lensing mass. In total, the additional travel time of a photon from the source to the observer due to the gravitational lensing effect of a mass, $M$, is given by \citep{1986ApJ...310..568B};
\begin{equation}
    c t = \frac{D_L D_S}{2 D_{LS}} \left( \beta - \theta \right)^2
    - \frac{4 G M}{c^2} ln \left| \theta \right|
\label{eqn:time}
\end{equation}
up to an additive constant 
term which is independent of $\beta$ and $\theta$.
Here, $D_L$ is the distance to the lens, $D_S$ is the distance to the source and $D_{LS}$ is the distance between the two.
Whilst the total additional travel times for an individual line-of-sight is not observable, the time-delay between images is a key observable and has been used as
a probe of cosmology on large scales~\citep{1964MNRAS.128..307R,2017MNRAS.468.2590S}. 

Again, following \citet{2019MNRAS.486.5400S}, we adopt the values of $D_L=4$kpc, $D_S=8$kpc, $M=0.08M_\odot$ and $\beta = \frac{1}{2}\theta_E$. With this, $\theta_E = 1.4\times10^{-9}$ radians, or $2.9\times10^{-4}$ arcseconds, and the image separation is $\sim2.1 \theta_E$. Given the symmetry of the lensing configuration, this is also the angle separating the lines of sight from the source star, a point that will become salient in Section~\ref{sec:decoherence}. 

\subsection{Photon Bunching and Microlensing}
\label{sec:photonbunch}
The HBT effect represents a correlation of intensity seen between two nearby lines-of-sight to a source
\citep{1956Natur.177...27B,1957RSPSA.242..300B,1958RSPSA.243..291B,1958RSPSA.248..199B,1958RSPSA.248..222B}.
At the quantum level, the HBT effect corresponds to a correlated bunching of photon counts between the two sight-lines~\citep{1961AmJPh..29..539F}, finding applications in the bunching bosons and fermions in studies of particle physics~\citep[e.g.][]{bosonarticle}.

\citet{2019MNRAS.486.5400S} considers the HBT effect during gravitational lensing, suggesting that the correlation of photon bunching between the two sight-lines to a source can be used to determine the resultant time delay. It is noted that the average time delay between the images in the assumed gravitational microlensing scenario (Section~\ref{sec:microlensing}) is $\Delta t_{lens}\sim 1.5 \mu s$. Additionally, the time delay across an extended, but still unresolved image is not uniform, with a gradient due to the differing sight-lines~\citep{2002MNRAS.334..905G,2003ApJ...594..107Y}; for the situation considered here, the scale of time delay dispersion across an image is of order $\sim 6ns$. 

To simulate the HBT effect, \citet{2019MNRAS.486.5400S} adopt a spectral emission from the source with the form of a Lorentzian profile given by
\begin{equation}
S(\nu) \propto \frac{1}{1 + \left( 2 \pi \Delta \tau (\nu - \nu_o) \right)^2 }
\label{eqn:lorentz}
\end{equation}
Here, $\Delta\tau$ is the coherence time and sets the spectral width of the profile; again, the assumed value of $\Delta\tau$ is $\sim 10^{-8}s$ Furthermore, one hundred points of emission are considered over the source, with waves given random phases, and \citet{2019MNRAS.486.5400S} suggests that the HBT effect does not depend upon the central frequency, and so sets $\nu_o=0$. As the images of the two gravitationally lensed sources are unresolved, the intensity at the observer is given by
\begin{equation}
I(t) = \left| \sqrt{A_1} E(t) + \sqrt{A_2} E(t + \Delta t_{lens} ) \right|^2
    \label{eqn:intensity}
\end{equation}
where $E(t)$ is the electric field from each of the images, $A_i$ is the magnification (Equation~\ref{eqn:magnification}); note that these quantities represent the sum over the one hundred sample points over the source.
As demonstrated in Figure~2 of \citet{2019MNRAS.486.5400S}, the resultant light curve shows substantial variations on sub-microsecond time scales, and, through an auto-correlation of the signal, the time delay between the images becomes apparent as distinct individual peaks, suggesting this approach is viable. However, it is noted that such observations are beyond the temporal sampling of the current technology and new approaches with 30m-class telescopes will be required.

\section{Impact of Decoherence}
\label{sec:decoherence}
The angular scale over which the radiation from a source will be observed to exhibit a high degree of coherence in an instrument is given by several well-known criteria in optics, such as the Rayleigh or Michelson resolution limits.
For a telescope of size (diameter/baseline) $B$ operating at wavelength $\lambda$, these criteria are both of order:
\begin{equation}
    \theta_{resolving~power} \sim \frac{ \lambda }{ B } = \frac{ D_*} {D_{v}}
    \label{eqn:resolving}
\end{equation}
where the second fraction is a geometrical expansion of the (very small) angle into to the source diameter $D_*$ and viewing distance $D_{v}$ to the celestial target. 
It can be seen from this equation that given a star of size $D_*$, an interferometer must sample the radiation field over a narrow range of angles or else coherence will vanish.
A trivial re-arrangement of Equation~\ref{eqn:resolving} gives this intrinsic angular coherence of the far field radiation:
\begin{equation}
    \theta_{coherent} \sim \frac{ \lambda }{ D_* } = \frac{ B } {D_{v}}
    \label{eqn:coherent}
\end{equation}
where instruments whose baselines subtend significantly larger angles to the source will lie beyond the criteria for coherence.

If we consider  $\lambda \sim 1 \mu m$ and $D_* \sim D_\odot$, this corresponds to $\theta_{coherent} \sim 10^{-15}$ radians or $\sim 3 \times 10^{-10}$ arcseconds.
So in order not to over-resolve the source, for an interferometer constructed at the plane of the lens at a distance $D_v = D_{LS} = 4$kpc away, the baseline $B$ should be less than about 100\,km or so. 
Unfortunately, this is six orders of magnitude smaller than the $\sim$AU separation of the light paths arriving at the lens plane from the source and subsequently redirected to the observer as described in Section~\ref{sec:microlensing}. 
This indicates that light traversing the two lines-of-sight from the source to observer by way of the lens will be completely incoherent: there will be no common embedded time signal for an auto-correlation to `latch' onto. 
Therefore we find that an application of the HBT effect at optical wavelengths will be  unable to yield the required microlensing time-delay.   

\begin{figure}
    \includegraphics[width=\columnwidth]{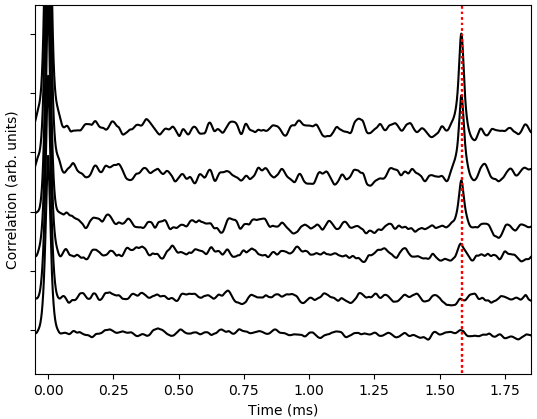}
        \caption{
        Recreation of the auto-correlation of the microlensing intensity as presented in \citet{2019MNRAS.486.5400S} but considering different central frequencies of the spectral emission distribution (Eqn.~\ref{eqn:lorentz}), going from $0 Hz$ (top line) to $12.5\times10^7 Hz$ (bottom line) in steps of $2.5\times10^7 Hz$; as can be seen, the peak due to the microlensing time delay (highlighted by the red dashed line) diminishes as higher central frequencies are considered.
        }
        \label{fig:cross}
\end{figure}

It is interesting to address why the simulations of \citet{2019MNRAS.486.5400S} resulted in the detection of auto-correlation peaks at $\Delta t_{lens}$ given the above angular scale of decoherence, $\theta_{coherent}$.
The flaw in the analysis  is the assumption that HBT effect is sensitive only to the width of emission spectrum, but not the central frequency, choosing $\nu_o=0 Hz$ in Equation~\ref{eqn:lorentz}.
By considering the frequency where the emission spectrum falls to a $10^{\rm th}$ of its peak value, 
this corresponds to a maximum frequency of waves considered in the emission from the source of
\begin{equation}
    \nu_{max} \sim \frac{ \sqrt{10} }{2\pi\Delta\tau}
    \label{eqn:freqmax}
\end{equation}
which, for $\Delta \tau = 10^{-8}s$, is $\nu_{max} \sim 5\times10^7 Hz$. This maximum frequency implies a minimum wavelength of $\lambda_{min} \sim 6m$; hence the waves emitted from the source included in the simulation of \citet{2019MNRAS.486.5400S} are typically metre-scale and above. 
Furthermore, if we consider a wavelength of $\lambda \sim 1m$, the angular resolving power (Equation~\ref{eqn:resolving}) over the AU-scale baseline provided by the geometry of the light-paths traversing the microlens system is $\theta_{resolving~power} \sim 7\times10^{-12}$ radians. 
It is instructive to note that this is similar to $1.1\times10^{-11}$ radians: the angle subtended by a star with the same diameter as the sun when viewed at a distance of 4\,kpc (the distance between lens and source). Hence, as the effective wavelengths considered by the previous study were of this scale and longer, the source object can therefore be considered to largely present as an unresolved point at these wavelengths. 
To further illustrate the point by recreating the microlensing analysis, but considering a range of central frequencies of the spectral energy distribution (Equation~\ref{eqn:lorentz}). Figure~\ref{fig:cross} presents the resultant auto-correlation signal, equivalent to the left-hand side of the lower-panel of Figure~2 in \citet{2019MNRAS.486.5400S}, for a central frequency of $\nu_o=0 Hz$ (top line) to $12.5\times10^7 Hz$ (lowest line) in steps of $2.5\times10^7 Hz$. The cross-correlation peak due to the gravitational microlensing time delay, highlighted by the dashed red line decreases in amplitude as high central frequencies are considered, vanishing into the noise by the highest central frequency is considered.  
It is therefore unsurprising that the HBT coherence computed between the two light paths produced a signal through which the auto-correlation can identify $\Delta t_{lens}$ and we concur that should radio observations of this type be feasible, then the expected signals should be present in the correlation.
However, this does demonstrate that for observations in the optical, which are currently the focus of microlensing surveys, decoherence between the light-paths means that measuring the lensing time-delay is beyond the reach of the HBT effect.    

\section{Conclusions}
\label{sec:conclusions}
In this paper, we have reexamined the claim of \citet{2019MNRAS.486.5400S} that the photon correlations from a thermal source (commonly identified as the HBT effect) can be used to measure the short time delay between multiply imaged components during a gravitational microlensing event. We find that the assumption that the HBT effect is independent of the observed central frequency is flawed, resulting in the de-facto adoption of metre-scale wavelengths in presented simulations of \citet{2019MNRAS.486.5400S}. Such signals can (in principle) indeed be correlated over the range of angles spanned by sight-lines traversing the system to generate the different lensed sub-images. However, this angular scale is many orders of magnitude larger than the coherence scale for optical wavelengths, demonstrating that measuring the gravitational microlensing time delays with the HBT effect with existing optical approaches is not possible. 

In closing, we note that the question of determining gravitational lens time delays through interference/correlation effects has a long history, and limitations due to source coherence are well known \citep[e.g.][]{1985A&A...148..369S}. However, when a source straddles a caustic, where the angular separation and time delay between adjacent light rays goes to zero, transient interference effects may be measurable, although signals might be swamped when integrating the decoherence over the entire source \citep{1991SvA....35...11M,1991SvA....35..116M,2003ApJ...594..456Z}. 
We also note claims, expanded over a series of papers, that different path lengths in gravitational lens systems might carry time separated correlations as an application of the \citet{1958AmJPh..26..481A} effect \cite[see][]{2008MNRAS.389..364B,2011MNRAS.411.1695B,2013MNRAS.436.1096B}. 
However identical reasoning to that already presented here for the scheme of \citet{2019MNRAS.486.5400S} removes the chance of such correlated signals, regardless of the configuration of the detection apparatus.

\section*{Acknowledgements}
GFL acknowledges support from a Partnership Collaboration Award between the University of Sydney and the University of Edinburgh, and appreciates the hospitality of the Royal Observatory, Edinburgh, where he finally understood intensity interferometry. PT acknowledges helpful discussions with William Tango. 




\bibliographystyle{mnras}
\bibliography{photon} 

\bsp	
\label{lastpage}
\end{document}